\documentclass[preprint,12pt]{elsarticle}
\usepackage{amssymb}
\journal{}

\begin{document}

\begin{frontmatter}

\title {Local external signal shows larger effectiveness when imposed on the fast components in an array of inductively coupled Josephson junctions}

\author{Alireza Valizadeh}

\address{ Institute for Advanced Studies in Basic Sciences, Zanjan 45195-1159, Iran}

\ead{valizade@iasbs.ac.ir}

\begin{abstract}
An array of inductively coupled Josephson junctions is considered and it is shown that in presence of a fast impurity which is a junction with a smaller value of critical current, a small amplitude periodic signal can be detected all elsewhere in the array, when is imposed on the fast component. On the other hand, small amplitude signals which are imposed on the other oscillators in the array have minor effect on the dynamics of the whole array.
\end{abstract}

\end{frontmatter}

Arrays of Josephson junction are prototype of the nonlinear systems with many degrees of freedom\cite{array1,array2}. While external periodic signals can entrain the dynamics of a single junction leaving plateaus in the current-voltage (I-V) characteristics of the junction\cite{shapiro}, interaction of the external signals with the coupled junctions in the arrays lead to more complex dynamics e.g. fractional Shapiro steps can be seen as the vortex lattice moves in accordance to the external signals in two-dimensional lattices\cite{fractional1,fractional2}.

Following a recent study\cite{vks2} here we investigate influence of a local periodic signal on the dynamics of a chain of linearly coupled Josephson junctions. The equations can be reduced to the well-known Frenkel-Kontorova (FK) model which its applications are from pedagogical system of mechanical transmission line consisting of linearly coupled pendula\cite{fkbook}, to dislocation dynamics in metals\cite{dislocation1,dislocation2}, DNA dynamics\cite{DNA} and strain waves in earthquakes\cite{quake}. A single {\it fast} impurity in the FK model can serve as the source of solitary waves and gets a leading role in the dynamics of the whole array regardless of the length of the array\cite{vks2}. Here we show such an arrangement shows different responses to the locally imposed external signals, depending on where the signal is imposed.

We consider a parallel array of Josephson junctions which are coupled inductively\cite{mccumber}

\begin{eqnarray}
 \frac{\hbar C_{j}}{2e}\ddot{\theta}_{j}
+\frac{\hbar}{2eR_{j}}
\dot{\theta}_{j}+I_{cj}\sin \theta_{j} =I_{j} + E_{j} \sin \omega t
+\frac{\Phi_{0}}{2\pi}[\frac{1}{L_{j}}
(\theta_{j+1}-\theta_{j})-\frac{1}{L_{j-1}} (\theta_{j}
-\theta_{j-1})],
\end{eqnarray}
where $C_{j}$, $R_{j}$ and $I_{cj}$ are the capacitance, resistance, and critical current of the $j$th junction, respectively; $\theta_j$ is the phase of the superconducting order parameter on the $j$th junction; and $L_{j}$ is the inductance of the $j$th plaquette. $\Phi_{0}=hc/2e$ is the flux quantum and $I_{j}$ and $E_{j}$ are the constant current and the amplitude of the periodic current of $j$th junction, respectively.

Scaling the parameters of the junctions by $C_{0}$, $R_{0}$ and $I_{c0}$, and the inductance by $L_{0}$, we can introduce a dimensionless time $\tau=\omega_{p} t$ where $\omega_{p}=\sqrt{2eI_{c0}/\hbar C_{0}}$ is the plasma frequency. With equal inductances Eq. (1) becomes
\begin{eqnarray}
\ddot{\theta}_{j} +  \beta_{c}^{-1/2}
\dot{\theta}_{j}+\alpha_{j} \sin \theta_{j} = i_{j} + \epsilon_{j}
\sin \omega \tau+ k_{0}(\theta_{j+1}-2
\theta_{j}+\theta_{j-1}).
\end{eqnarray}
Here $\beta_{c}=2e R_{0}^{2}I_{c0}C_{0}/\hbar$ is the McCumber parameter, which characterizes the relative importance of the damping, and $k_{0}=\Phi_{0}/2\pi I_{c0}L_{0}$ is the linear coupling constant. The normalized critical current of the junctions is $\alpha_{j}={I_{cj}}/{I_{c0}}$, $\omega$ is the drive frequency rescaled by the plasma frequency, and over-dots indicate derivatives with respect to $\tau$.

When the $i_{j}$ is small, the $\theta_{j}$ may oscillate but remain bounded and the average voltage across each junction is zero. For sufficiently large $i_{j}$, the $\theta_{j}$ increases with time with an average rate of increase that is proportional to the voltage difference across the junction. Then we will say that the junctions are rotating. With inductive coupling in the steady state, the average voltage across all the junctions in the array, and so the average rate of change of the phases, would be equal. Hence I-V characteristic of all the junctions in the array, that is the average voltage of the junction respect to the constant input current, is the same for all the junctions.

Equation 2 is a damped driven FK model\cite{fkbook}. We consider an almost homogeneous array into which has been introduced a junction (the impurity) that has a critical current relatively lower than that of the other junctions which in isolation tends to rotate with a larger frequency. All the junctions then are driven by a constant current which is larger than the critical current of the junctions, i. e. in isolation all the junctions would be in rotating state; the impurity rotates faster. Now we impose the periodic drive on the $m$th junction and change $m$ to see how the position of the locally imposed time-dependent signal changes the response of the whole system.

\begin{figure}[ht!]
\centerline{\includegraphics[width=13cm]{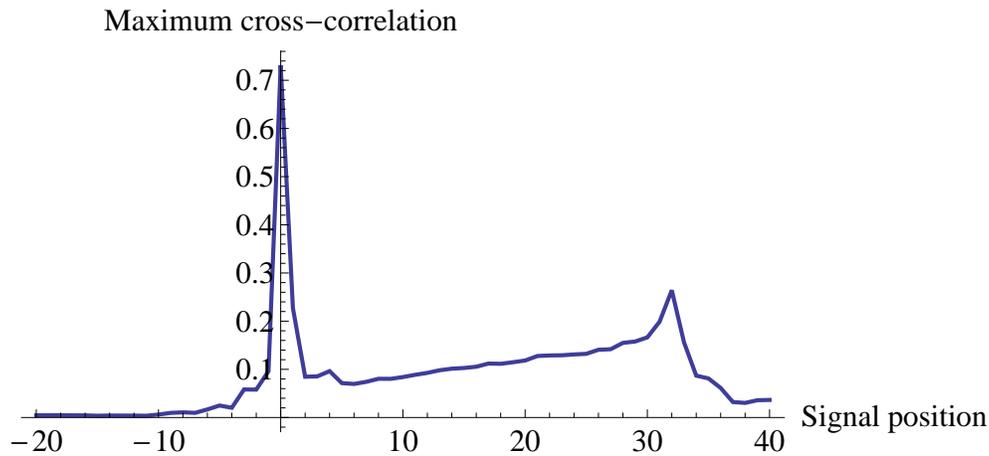}}
\vspace{0.1cm} \caption{Maximum cross-correlation of the derivative of a sample junction with external signal vs. position of the external signal i. e. the number of junction on which the external periodic signal is imposed. Total number of junctions is $127$ and the middle junction is labeled by $j=0$. Sample junction here is chosen $j=32$ but qualitatively, the result is independent of the number of sample junction while it is not so close to the impurity and to the boundary. Critical currents of all the junctions are chosen from a uniform distribution in the range $[0.99,1.01]$, except for the middle junction which is $\alpha_{0}=0.5$. External dc current is $i_{j}=1.01$ for all $j$. Damping parameter and coupling constant are $\beta_{c}^{-1/2}=0.75$ and $k_{0}=0.5$, respectively. The external periodic input with amplitude $a_{m}=0.4$ and frequency $\omega =0.5$ is imposed only on the $m$th junction and $m$ is varied from $-20$ to $40$. Boundary conditions are absorbing i. e. the dc drive of boundary junctions is switched off to prevent the waves to re-enter into the array.} \vspace{-0.0cm}
\label{fig1}
\end{figure}

In a long chain consisted of $127$ junctions, the fast impurity is located in the middle which is labeled by $j=0$. Using the standard definition of the cross-correlation function $\rho(T)$ of two signals $x_{1}$ and $x_{2}$
 \begin{equation}\label{eq3}
    \rho_{12}(T)=\frac{\langle x_{1}(t) x_{2}(t-T)\rangle-\langle x_{1}(t) \rangle \langle x_{2}(t-T)\rangle}{\sqrt{Var(x_{1})}\sqrt{Var(x_{2})}},
\end{equation}
with $\langle \rangle$ standing for time average and $Var$ for the variance, we calculate the cross-correlation of the derivative of the phase of a sample junction ($32$th junction in this manuscript) with the periodic signal. The cross correlation shows a maximum in a nonzero time lag which is dependent to distance of the sample junction with the impurity. We then change the junction on which the periodic signal is imposed and record maximum value of the cross-correlation as a measure of how the junction is influenced by the external signal. The result for typical values of the parameters is shown in Fig. 1. The plot shows the influence of the signal in the array is considerably boosted when it is imposed on the fast impurity. The response also shows a trivial maximum when the position of the signal approaches to the junction which we are studying.

\begin{figure}[ht!]
\vspace{-1cm}
\centerline{\includegraphics[width=12cm]{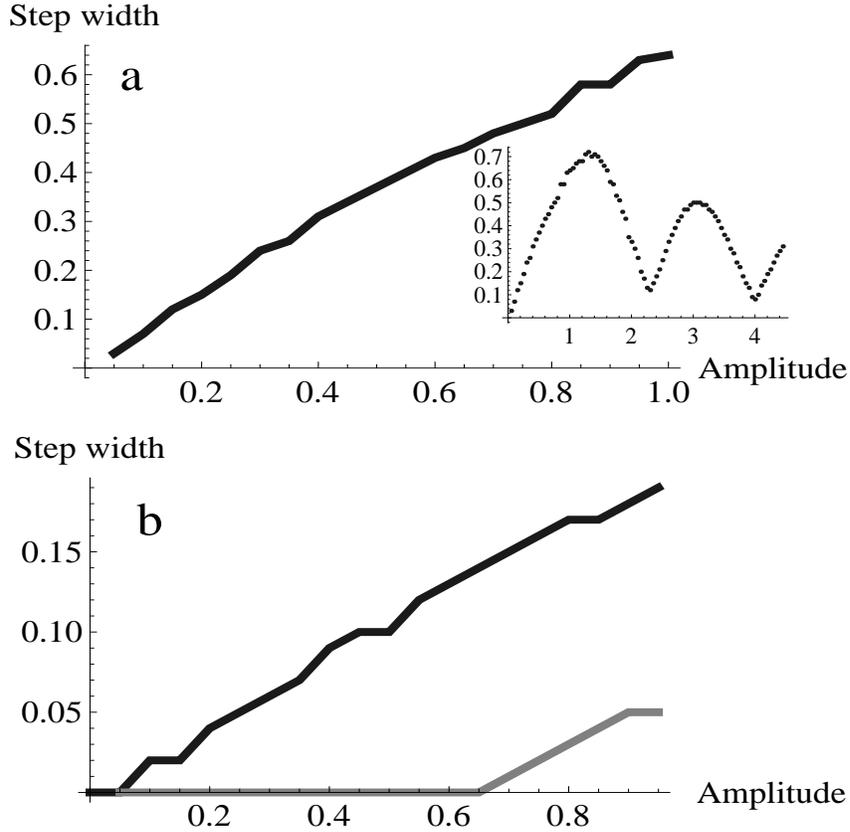}}
\vspace{-5.5cm} \caption{(a) The width of the main $1:1$ Shapiro step for a single junction as a function of the amplitude of periodic signal. (b) The width of main $1:1$ Shapiro step for an array when the local signal is imposed on the fast impurity (dark gray) and on the $10$th junction (light gray). Total number of junctions is $127$ and the middle junction is labeled by $j=0$. Sample junction here is chosen $j=32$ but the results are independent of the number of sample junction. Critical currents of all the junctions are chosen from a uniform distribution in the range $[0.99,1.01]$, except for the middle junction which is $\alpha_{0}=0.5$. External dc current is $i_{j}=1.01$ for all $j$. Damping parameter and coupling constant are $\beta_{c}^{-1/2}=0.75$ and $k_{0}=0.5$, respectively. } \vspace{-0.0cm}
\label{fig4}
\end{figure}

The role of the fast impurity to create a {\it position sensitive response} to the local inputs can also be revealed by studying I-V characteristic of the system. For the single junctions it is known that the width of the Shapiro steps shows a Bessel function type relation to the amplitude of the signal as it is shown in Fig. 2a\cite{bessel}, and for small amplitude signals which we consider in this study, the width of the main Shapiro step has a linear relation with the amplitude of the periodic current. As noted before, in the array with inductive couplings, I-V characteristic for all the junctions would be the same; if the transient state is left when integrating the voltage. We have shown the width of the main Shapiro step in the characteristic of a sample junction in the array when the signal is imposed on the impurity and also on another junction in the array (Fig. 2b). It can be seen the width of the main step for the junctions in the array shows a linear relation with signal amplitude, when signal is imposed on the impurity. On the other hand, if the signal is located on another junction, no locking region is seen for small amplitude signals.

Figure 2 also shows the off-impurity signal can entrain the array dynamics if its amplitude is larger than a critical value. This result interestingly confirms the role of the fast component: for off-impurity signals with amplitude larger than a threshold (about 0.7), the critical current of the junction which receives the signal decreases so that the impurity looses its role as the fastest junction (the junction with largest value of voltage when isolated). So the junction on which the signal is imposed can turn to the fastest junction and take the role of the leading junction. In this case entrainment of the array by a large amplitude off-impurity signal would be possible.

 Dependence of the response of the model to the frequency shows also interesting features. In Fig. 3 the maximum cross-correlation of the derivative of the phase of the junction with the periodic input signal, is plotted for a range of the frequencies of periodic signal, with the other parameters the same as those of the Fig. 1. Again two cases are considered: when the signal is imposed on the fast impurity and when it is placed on another junction. The figure shows a resonant effect for certain values of frequencies (about multiples of 0.6 for both cases), but the resonance amplitude is considerably larger when the signal is located on the fast impurity. Note that since the critical currents of the impurity and the other junctions are different, in isolation they would have different resonant frequencies. So one can take the result of Fig. 1 just because of the external frequency is close to the resonance frequency of the impurity. Figure 3 rejects this possibility: whether the external signal is imposed on the impurity or on the other junctions, the resonant frequency would be identical and for all the values of frequencies the response of the array is larger if the signal is imposed on the impurity.

\begin{figure}[ht!]
\vspace{-0cm}
\centerline{\includegraphics[width=12cm]{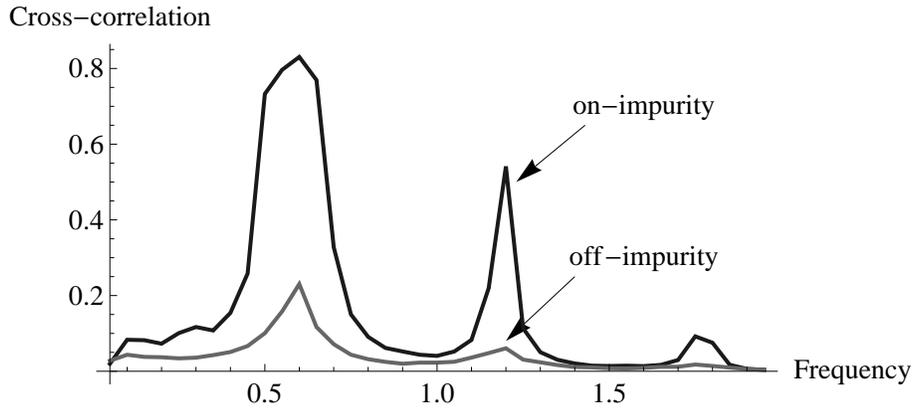}}
\vspace{-0cm} \caption{(a) Maximum cross-correlation of the derivative of a sample junction with external signal vs. frequency of the external signal when the local periodic input with amplitude $a_{m}=0.4$ is imposed on the impurity (dark) and on $10$th junction (light gray).  Total number of junctions is $127$ and the middle junction is labeled by $j=0$. Sample junction here is chosen $j=32$ but qualitatively, the result is independent of the number of sample junction while it is not so close to the impurity and to the boundary. Critical currents of all the junctions are chosen from a uniform distribution from the range $[0.99,1.01]$, except for the middle junction which has smaller critical current $\alpha_{0}=0.5$. External dc current is $i_{j}=1.01$ for all $j$. Damping parameter and coupling constant are $\beta_{c}^{-1/2}=0.75$ and $k_{0}=0.5$, respectively. } \vspace{-0.0cm}
\label{fig4}
\end{figure}

 The role of the fast component in generating solitary pulses consisting of a pair of kink-antikink, can explain all the above results. Imposing the periodic signal on the fast component, the rate of the nucleation of the kinks which is proportional to the rate of the change of the phase of the junction, can be locked to the external frequency. The solitary pulses themselves move along the array and entrain the whole array after finite transient time which grows with the size of the array. A signal when imposed on the other junctions, can entrain the junctions in the vicinity of where the local signal is imposed but has no effect on the rate of solitary pulses which are yet produced by the impurity. So a long range influence is not expected by off-impurity signals. We also note that the feature of the FK model in sporting solitary excitations, allows a behavior like what is seen above: a ladder arrangement of Josephson junction is a counterexample in which the long range effect of local signals can not be seen in it.

\end{document}